\documentclass[twocolumn,superscriptaddress,english,pre,longbibliography]{revtex4-2}

\usepackage{titlesec}
\renewcommand{\thesection}{\arabic{section}} 
\renewcommand{\thesubsection}{\thesection.\arabic{subsection}} 
\renewcommand{\thesubsubsection}{\thesubsection.\arabic{subsubsection}} 

\titleformat{\section}
  {\normalfont\Large\bfseries}
  {\thesection.}
  {1em}
  {}

\titleformat{\subsection}
  {\normalfont\normalsize\bfseries}
  {\thesubsection.}
  {1em}
  {}

\titleformat{\subsubsection}
  {\normalfont\normalsize\bfseries}
  {\thesubsubsection.}
  {1em}
  {}

\usepackage{listings}
\usepackage{xcolor}

\definecolor{codebg}{rgb}{0.95,0.95,0.95} 
\definecolor{codecomment}{rgb}{0,0.6,0} 
\definecolor{codestring}{rgb}{0.58,0,0.82} 
\definecolor{codekeyword}{rgb}{0,0,1} 

\usepackage[colorlinks=true,urlcolor=blue,citecolor=blue,linkcolor=blue]{hyperref}
\usepackage{makecell}
\usepackage{enumitem} 
\usepackage{graphicx}
\usepackage{dcolumn}
\usepackage{bm}
\usepackage{amsmath,amssymb,amsthm,mathrsfs,amsfonts,dsfont}
\usepackage{balance}
\usepackage{color}
\usepackage{array}
\usepackage{tabularx}
\usepackage{longtable}
\usepackage{color}
\usepackage{float}
\usepackage{indentfirst}
\usepackage{txfonts}
\usepackage{booktabs} 
\usepackage{multirow}

\usepackage{tocvsec2}
\usepackage{placeins}

\usepackage{physics}
\usepackage{algorithm} 
\usepackage{algorithmic}
\usepackage{subfloat}
\usepackage{graphicx}

\usepackage[all]{hypcap}
\makeatletter
\let\saved@includegraphics\includegraphics
\AtBeginDocument{\let\includegraphics\saved@includegraphics}
\makeatother
\usepackage{braket}
\usepackage{tikz}
\usepackage{float}
\usepackage{amsmath,amssymb,amsthm,mathrsfs,amsfonts,dsfont}
\usepackage{xcolor}

\begin{document}

\title{Reconstruction of detector error model for quantum error correction}

\author{
Cheng Ye
}
\affiliation{CAS Key Laboratory for Theoretical Physics, Institute of Theoretical Physics, Chinese Academy of Sciences, Beijing 100190, China}
\affiliation{School of Physical Sciences, University of Chinese Academy of Sciences, Beijing 100049, China}

\author{
Pan Zhang
}
\email{panzhang@itp.ac.cn}
\affiliation{CAS Key Laboratory for Theoretical Physics, Institute of Theoretical Physics, Chinese Academy of Sciences, Beijing 100190, China}
\affiliation{School of Physical Sciences, University of Chinese Academy of Sciences, Beijing 100049, China}
\affiliation{School of Fundamental Physics and Mathematical Sciences, Hangzhou Institute for Advanced Study, UCAS, Hangzhou 310024, China}
\affiliation{Beijing Academy of Quantum Information Sciences, Beijing 100193, China}

\date{June 15, 2026}

\begin{abstract}
Fault-tolerant quantum computing fundamentally relies on the accurate characterization of circuit-level noise to optimize decoding algorithms. However, extracting complex multi-body error correlations remains challenging. Contemporary greedy inference algorithms can suffer from statistical distortion, discarding true physical mechanisms while introducing many unphysical false positives. Here, we introduce the Correlation-Analysis-based Hypergraph Reconstruction (CAHR) algorithm, a globally consistent framework to invert experimental syndrome statistics directly into discrete physical hypergraphs. By coupling exact algebraic correlation equations with a top-down concurrent-pruning strategy, CAHR recovers the fault topology without false positives for both $d=5$ rotated surface codes and dense 8-body 2D color codes in our benchmark settings. Furthermore, we show that exact continuous parameter extraction in dense codes is limited by a \textit{variance cascade}, where absolute statistical variance accumulates linearly from high- to low-degree mechanisms. This motivates a two-stage inference paradigm: utilizing CAHR to extract the fault topology, followed by continuous probability optimization. This provides a practical approach for characterizing and decoding highly correlated noise in realistic quantum hardware.

\end{abstract}

\maketitle

\section{Introduction}
The realization of fault-tolerant quantum computing relies on Quantum Error Correction (QEC) to actively protect quantum information from environmental decoherence and operational imperfections \cite{Nielsen2010, Terhal2015, Campbell2017}. A central bottleneck in this pursuit is the precise characterization of complex, circuit-level noise to provide high-quality prior information for decoding algorithms \cite{Spitz2018AdaptiveWE}. To formalize these intricate dynamics, the Detector Error Model (DEM) provides a concise probabilistic hypergraph representation \cite{Gidney2021}. While standard two-point graph representations suffice for simple noise, realistic extraction protocols and advanced topological codes—such as surface codes \cite{Dennis2002, Fowler2012} and 2D color codes \cite{Bombin2006}—inherently generate complex, multi-body error correlations \cite{Harper2020}. These manifest as high-order hyperedges within the DEM, representing single physical faults that simultaneously trigger multiple detectors \cite{Takou2025}.

Systematically reconstructing arbitrary-order hypergraphs directly from experimental syndrome statistics remains a formidable challenge. Recent works have already established exact correlation equations and algebraic inversion routes for DEM estimation \cite{Remm2026Experimentally, BlumeKohout2025EstimatingDE, Arms2025EstimatingDE}. Among them, \cite{Arms2025EstimatingDE} provides the most direct empirical reference for DEM reconstruction, while \cite{Remm2026Experimentally, BlumeKohout2025EstimatingDE} primarily supply on correlation analysis and algebraic inversion. However, practical topology discovery still faces a severe combinatorial bottleneck. Existing workflows often rely on order-by-order (greedy) inference or computationally expensive maximum-likelihood optimization, and sequential greedy paradigms are particularly vulnerable to finite-sample contamination: when lower-order correlations are fitted before higher-order mechanisms are resolved, unmodeled high-order physical errors pollute the statistical estimators, distort residual correlations, and can simultaneously generate false positives and suppress genuine hyperedges. Distinct from these topological failures, extracting exact continuous probabilities is further limited by a \textit{variance cascade}—a phenomenon where statistical variance accumulates from high-order parent hyperedges down to low-order child faults, making precise probability calibration increasingly fragile in dense hypergraphs.

In this work, we address this topology-discovery bottleneck by introducing the Correlation-Analysis-based Hypergraph Reconstruction (CAHR) algorithm.
The mathematical foundation of CAHR relies on the exact analytical equations of arbitrary-order syndrome correlations that have already been derived in the recent literature \cite{Remm2026Experimentally, BlumeKohout2025EstimatingDE, Arms2025EstimatingDE}. CAHR is a top-down workflow that combines permissive candidate generation with concurrent pruning, allowing the discrete hypergraph topology to be identified before pursuing high-accuracy continuous parameter fitting. An alternative derivation of the same correlation equations within a Tensor Network Detector Error Model (TNDEM) formalism is provided in the Appendix where TNDEM serves as a structured derivational framework that interfaces naturally with tensor-network contraction.

We validate CAHR on both $d=5$ rotated surface codes and the denser 8-body $d=5$ 2D color codes. The latter is useful because directly comparable DEM-reconstruction results for dense color-code settings remain scarce. Within these benchmarks, CAHR attains exact topology reconstruction with zero false at $5 \times 10^6$ and $1.5 \times 10^7$ shots, respectively. We further show that supplying the reconstructed topology to Belief Propagation with Ordered Statistic Decoding (BP-OSD) \cite{Panteleev2021, Roffe2020} restores convergence toward the ideal logical error rate (LER).

By isolating discrete geometric structure from cascaded probability errors, we therefore advocate a two-stage decoupled inference paradigm. In this view, CAHR is best understood as a Stage-1 topology-discovery tool that resolves the discrete hypergraph, while high-accuracy continuous parameter refinement can be delegated to downstream statistical or learning-based methods on the fixed topology \cite{Spitz2018AdaptiveWE, Wagner2023Learning, Sivak2025RL}. This division of labor is useful in dense correlated-noise settings, where topology identification and parameter calibration exhibit different statistical difficulties.

The paper is organized as follows. Section II details the our algorithm construction, derives the exact analytical relationships, and introduces the top-down double-pruning algorithm. Section III investigates the impact of finite-sample fluctuations and the variance cascade, demonstrating end-to-end LER validation on distinct topological codes. Finally, Section IV discusses the physical implications of our findings and formally outlines the two-stage decoupled inference paradigm.

\section{Methodology}
In realistic fault-tolerant quantum computing, conventional qubit-level Pauli error models are often insufficient to capture the intricate, correlated fault dynamics introduced by noisy syndrome extraction circuits. To accurately characterize these processes, the Detector Error Model (DEM) is widely adopted \cite{Gidney2021, Higgott2025SparseBlossom, mcewen2023relaxing}. A DEM formalizes the probabilistic relationship between physical error mechanisms and the triggered detectors across the spacetime circuit. Mathematically, a DEM is represented as a hypergraph $G=(V,E)$, where the vertex set $V$ corresponds to the detectors, and a hyperedge $e$ in set $E$ represents independent physical error mechanisms that simultaneously flip all detectors in set $e$.

\subsection{Analytical Solution for Arbitrary-Order Correlations}
Building upon a Tensor Network (TN) formalism of the DEM (with the complete derivation provided in the Appendix), we establish an exact algebraic mapping between experimental observables and physical error probabilities. We define the observable variable $\sigma_i = 1 - 2s_i$, where $s_i \in \{0,1\}$ is the binary syndrome of detector $i$. The correlation of a subset of detectors $\alpha$ is governed by the moment equation:
\begin{equation}\label{eq:moment_structure}
    M_{\alpha} = \langle \prod_{i\in \alpha} \sigma_i \rangle = \prod_{|e \cap \alpha| \equiv 1 \pmod 2}(1-2p_e).
\end{equation}
This system consists of $2^{|V|}-1$ equations corresponding to all possible error mechanisms. Rather than relying on logarithmic transformations used in prior heuristic approaches, we preserve the multiplicative structure and introduce a set of recursive helper functions, $f_{\beta}$, to achieve a compact, closed-form inversion:
\begin{equation}\label{eq:f-m}
    f_{\beta} = \prod_{\alpha\subseteq \beta} M_{\alpha}^{(-1)^{|\alpha|-1}}.
\end{equation}
Using these helper functions, the exact probability of an arbitrary-order hyperedge $\alpha$ is analytically resolved as:
\begin{equation}\label{eq:p-f}
    p_\alpha = \frac{1}{2}\left[ 1 - \frac{\left(f_\alpha \right)^{1/C_{\alpha}}}{\prod_{\beta \supsetneq \alpha} (1-2p_{\beta})} \right],
\end{equation}
where the scaling exponent is $C_{\alpha} = 2^{|\alpha|-1}$. This closed-form solution is exact for arbitrary-order hyperedges within this formalism, avoiding the need for complex optimization loops and the associated local-minimum issues of numerical fitting procedures.

It is important to note that equivalent systems of equations and their algebraic inversions have already been derived in several recent works \cite{Remm2026Experimentally, BlumeKohout2025EstimatingDE, Arms2025EstimatingDE}. Our TN-based treatment is used here as an alternative structured derivation, which is convenient for organizing the correlation equations and connecting them to tensor-network contraction.

\subsection{Inference with Finite Statistics and Structural Priors}

While the original full-rank system yields a unique solution corresponding to the ideal case, practical QEC circuits are constrained by finite sample sizes ($N_{\text{shots}}$) and possess hypergraphs of limited orders where the physical error rates of non-existent hyperedges are zero. Finite sampling introduces statistical fluctuations, causing the analytical solutions for these non-existent edges to manifest as small, non-zero values. 

To address this discrepancy, we modify the solution to maintain consistency with a given topological prior (a candidate hypergraph set $E$). We compute only the necessary equations for existing edges $\alpha \in E$ and remove the terms of non-existing edges in the denominator by enforcing a zero probability:
\begin{equation}\label{eq:p-f-given}
    p_{\alpha} = \frac{1}{2} - \frac{1}{2} \frac{\left(f_{\alpha} \right)^{1/C_{\alpha}}}{\prod_{\beta \supsetneq \alpha, \beta \in E} (1-2p_{\beta})}.
\end{equation}

Note that the helper function still consists of $M_{\alpha}$ for all $\alpha$ whether in $E$ or not. In previous work, the instance constraining the correlation in pair order has been proved to be consistent with the classical pair correlation analysis\cite{Spitz2018AdaptiveWE}.

\subsection{Correlation-Analysis-based Hypergraph Reconstruction}

In experimental scenarios, the true set of active physical error mechanisms $E$ is \textit{a priori} unknown and must be directly inferred from the measured syndrome data. To systematically reconstruct the discrete hypergraph and suppress the combinatorial explosion of high-order candidates without human bias, we propose Correlation-Analysis-based Hypergraph Reconstruction (CAHR) algorithm, via global necessary-condition search and concurrent pruning.

Before detailing the procedure, it is important to clarify our departure from the sequential, order-by-order greedy growth strategies prevalent in recent literature. While the theoretical foundation of greedy hyperedge growth is mathematically rigorous, its layer-by-layer statistical filtering is ultimately excessive for our primary objective. The fundamental bottleneck in arbitrary-order correlation analysis is the combinatorial explosion of potential hyperedges, which scales strongly with the number of detectors $|V|$ and the hyperedge degree $k$. By enforcing the simple geometric necessary condition of pair-correlation cliques under a reasonable finite-sample regime (where initial false-positive pairs are sufficiently suppressed), we already achieve our core goal: compressing the large combinatorial search space down to a computationally tractable candidate pool. At this stage, further heuristic structural filtering offers only marginal benefits.

Furthermore, given the complex and often unpredictable propagation of finite-sample statistical noise across different correlation orders, we argue that the explicitly calculated physical error probability, $p_{\alpha}$, serves as the most reliable metric for hyperedge validation within this framework. As dictated by our exact analytical solution in Eq.~\eqref{eq:p-f-given}, the accurate computation of $p_{\alpha}$ requires the prior knowledge of all its parent hyperedges ($\beta \supsetneq \alpha$). This algebraic dependency necessitates a top-down evaluation sequence. Consequently, our framework deliberately adopts a permissive candidate generation strategy—allowing all potential hyperedges identified by the pair-clique prior to exist initially—and defers the elimination of fluctuation-induced false positives to a \textit{concurrent} pruning executed during the global, top-down probability calculation. The complete reconstruction protocol proceeds as follows:

\begin{enumerate}
    \item \textbf{Pair-Correlation Analysis with Initial Pruning:} If high-order hyperedges exist in the underlying DEM, they inevitably project onto the pair-correlation space, yielding an apparent 2-body probability $\tilde{p}_{ij} = \frac{1}{2} - \frac{1}{2}\sqrt{f_{ij}}$. In the physically relevant regime ($p \ll 1$), this mapping is approximated as $\tilde{p}_{ij} \approx p_{ij} + \sum_{\{i,j\} \subsetneq \alpha} p_{\alpha}$. This leads to a simple geometric necessary condition: any valid high-order hyperedge $\alpha$ must project onto a fully connected clique of positive apparent correlations among its sub-pairs ($\tilde{p}_{ij} > 0$). However, finite-sample statistical fluctuations inevitably generate false-positive pair correlations. Without pruning, these noise-induced edges can interlink and produce a large number of unphysical ``ghost cliques.'' To suppress this effect, we calculate the pair correlations directly from the data and apply a conservative initial pruning threshold $\epsilon_{\text{pair}}$, removing any edge $(i,j)$ for which $\tilde{p}_{ij} < \epsilon_{\text{pair}}$.

    \item \textbf{Maximal Plausible DEM Formulation:} Treating the pruned $\tilde{p}_{ij}$ elements as an adjacency matrix, we then identify fully connected cliques up to a predefined maximum topological order ($k_{\max}$) using standard network-processing algorithms. Here $k_{\max}$ is chosen from prior knowledge of the syndrome-extraction circuit and code structure, namely the highest detector-correlation order that a single fault mechanism can generate. For example, we use $k_{\max}=4$ for the rotated surface code and $k_{\max}=8$ for the 2D color code. In settings with less prior knowledge, one may choose a more conservative larger $k_{\max}$, at the cost of a larger candidate space and more expensive clique search. The global candidate hypergraph $G_{\text{candidate}}$ is then constructed by including all sub-hyperedges contained within these bounded cliques. This structurally redundant yet geometrically constrained topology rules out regions where high-order hyperedges cannot exist, thereby reducing the combinatorial search space.

    \item \textbf{Global Correlation Analysis with Instant Pruning:} Finally, we execute the modified global correlation analysis over this candidate hypergraph $G_{\text{candidate}}$. The deliberately loose and redundant conditions of the previous step preserve topological completeness, but they also introduce many non-physical candidate hyperedges. Due to statistical fluctuations, the exact analytical inversion may assign small positive values to such dummy hyperedges, thereby producing false positives. We therefore apply a global pruning rule: immediately after calculating $p_{\alpha}$, the hyperedge is discarded if $p_{\alpha} < \epsilon_{\text{global}}$. As a result, finite-sample pruning inevitably carries a false-negative risk whenever the inferred probability of a genuine mechanism is pushed below zero or below the chosen non-negative threshold.
\end{enumerate}

Two remarks are important for interpreting this procedure. First, because the pair signal in Step 1 represents the projection of higher-order mechanisms, the initial screening should follow the signal scale of the higher-order mechanisms rather than that of genuine pair mechanisms. In the current implementation, if the projection of a genuine hyperedge is pruned at this stage, then that hyperedge will still be removed in Step 3 even if it is retained in the candidate topology.

Second, statistical fluctuations affect not only dummy hyperedges but also genuine weak mechanisms. In addition to producing small positive probabilities for non-physical candidates, finite sampling may also drive the inferred probabilities of genuine weak mechanisms to negative values when the sample size is insufficient. Since physical error rates require non-negative probabilities, only positive pruning thresholds are meaningful in practice.

The operational significance of integrating this pruning directly into the recursive evaluation process is that it maintains algebraic consistency. By resolving the analytical equations in descending order of hyperedge degree, this \textit{concurrent} truncation ensures that once a spurious high-order hyperedge is identified and pruned, its unphysical probability contribution is set to zero before it can propagate further down the hierarchy. In this way, the procedure decouples statistical artifacts from the underlying equation system and mitigates the downward propagation of finite-sample noise into the probability estimates of lower-order constituent edges.

\section{Results}

\subsection{Global Topological Reconstruction via Correlation Analysis}

\begin{table*}[!htb]
\centering
\begin{tabular}{@{}l c c c c c c@{}}
\toprule
\textbf{Topological Code} & \textbf{Analysis Shots} & \textbf{Total Edges ($|E|$)} & \textbf{False Pos.} & \textbf{False Neg.} & \textbf{CPU Time (s)} & \textbf{GPU Time (s)}\\
\midrule
\multirow{3}{*}{\textbf{Rotated Surface ($d=5, r=5$)}} 
& $5 \times 10^4$ & 5262 & 3944 & 361 & 8.76 & 4.01\\
& $5 \times 10^5$ & 2219 & 542  & 2   & 4.73 & 0.54\\
& $5 \times 10^6$ & 1679 & 0    & 0   & 7.37 & 3.17\\
\midrule
\multirow{3}{*}{\textbf{2D Color ($d=5, r=4$)}}
& $1.5\times 10^5$ & 1106 & 422  & 159 & 2.97 & 2.97\\
& $1.5\times 10^6$ & 860  & 31   & 14  & 1.26 & 1.26\\
& $1.5\times 10^7$ & 843  & 0    & 0   & 7.79 & 7.79\\
\bottomrule
\end{tabular}

\caption{\textbf{Exact Global Hypergraph Reconstruction via DEM Inference.} 
This table demonstrates the performance of our algorithm on a $d=5, r=5$ rotated surface code and a highly dense $d=5, r=4$ 2D color code (both at a physical error rate of $p=0.001$) across varying finite-sample regimes. To accurately isolate genuine correlations from finite-sample statistical noise, the topologies are evaluated using mechanism-specific pruning thresholds (e.g., $\epsilon_{\text{multi}} = 3 \times 10^{-5}$ for multi-detector correlations), anchored at $\approx p_{\min}/2$ of their respective error classes. For the surface code, finite-sample fluctuations are sufficiently suppressed at $5 \times  10^6$ shots, causing both false-positive and false-negative artifacts to vanish and recovering the ideal topology. For the highly dense 2D color code, which features hyperedges up to degree 8, the same criterion is reached at $1.5 \times 10^7$ shots. The execution time is decoupled into CPU core time (for graph generation and clique search) and GPU time (for batched tensor inversion).}
\label{tab:master_reconstruction}
\end{table*}

To establish the structural exactness of our framework, we first evaluate its capacity to reconstruct the complete physical hypergraphs of distinct topological codes from raw experimental statistics, which are efficiently generated using the high-performance stabilizer simulator Stim~\cite{Gidney2021}. Table~\ref{tab:master_reconstruction} summarizes the topological discovery performance. In these experiments, rather than applying a single uniform cutoff, we implement a mechanism-specific dual-threshold strategy: $\epsilon_{\text{single}}$ for single-detector errors (boundary edges) and $\epsilon_{\text{multi}}$ for multi-detector errors (correlating hyperedges). This distinction is critical because the anticipated physical error rates inherently differ across these two fundamental classes of fault mechanisms.

Grounded in statistical signal detection theory, both thresholds are set to approximately half of their respective minimum anticipated rates ($p_{\min}^{\text{single}}/2$ and $p_{\min}^{\text{multi}}/2$). Under finite-sample conditions, the inferred correlation $\tilde{p}_{\alpha}$ behaves as a random variable sampled from a Gaussian-like distribution centered around its true underlying value, $p_{\alpha}^{\text{Ideal}}$. Because $p_{\alpha}^{\text{Ideal}}$ is constrained to be either $0$ (false edges) or a genuine probability $p \ge p_{\min}$ (true physical error mechanisms), the midpoint $\epsilon \approx p_{\min}/2$ serves as a natural maximum-likelihood decision boundary for each respective error class when the two local hypotheses have comparable finite-sample widths. 

This practical choice is further informed by an explicit threshold-sensitivity check. In addition to the working choice $\epsilon \approx p_{\min}/2$, we also tested the lower and higher cutoffs $p_{\min}/4$ and $3p_{\min}/4$ for both the rotated surface-code and color-code benchmarks; the detailed results are summarized in Tables~\ref{tab:threshold_surface} and \ref{tab:threshold_color}. These data show that exact reconstruction is not tied to a single finely tuned threshold. At finite statistics, $p_{\min}/4$ admits more false positives, whereas $3p_{\min}/4$ generally gives slightly better overall results than $p_{\min}/2$. The reason is that the rule $p_{\min}/2$ is anchored to the weakest physical mechanism; for most hyperedges, this choice is therefore conservative. Nevertheless, in the main text we retain $p_{\min}/2$ as the reference threshold because it provides a simple, uniform prescription across the full candidate set.

Applying these optimal boundaries to the $d=5, r=5$ rotated surface code (Table~\ref{tab:master_reconstruction}, upper half) under a physical error rate (PER) of $p = 0.001$, we observe the direct impact of finite-sample suppression. The inherently sparse and localized nature of the surface code's syndrome extraction circuits yields a manageable combinatorial space. With lower statistics ($5 \times 10^4$ and $5 \times 10^5$ shots), statistical noise distributions are broad, inevitably causing the noise floor to cross the decision boundaries and admit false positives. However, by scaling the analysis to $5 \times 10^6$ shots, the noise variance is sufficiently compressed. Consequently, our method successfully infers the exact hypergraph topology without false positives, and in this benchmark the $50\%$ safety margin is sufficient to avoid truncating genuine weak mechanisms.

Pushing the algorithm to more demanding structures, we analyze the $d=5, r=4$ 2D color code at the same PER (Table~\ref{tab:master_reconstruction}, lower half). The lowest intrinsic probability of the true multi-detector physical hyperedges in the color code is of the same order of magnitude as that in the surface code, allowing us to utilize the identical threshold $\epsilon_{\text{multi}} = 3 \times 10^{-5}$. However, its highly non-local extraction circuits generate a dense entanglement structure featuring up to 8-body hyperedges. Due to this combinatorial growth, the resulting baseline statistical noise is significantly elevated. Separating this 8-body topology from the noise floor requires three times the sample volume. It requires $1.5 \times 10^7$ shots (compared against the $1.5 \times 10^5$ and $1.5 \times 10^6$ baselines) to ensure that the noises causing false-positive edges and false-negative errors are compressed below the decision boundaries, yielding accurate topological reconstruction. For context, color-code correlation-analysis studies have reported sample counts as high as $5 \times 10^7$ \cite{Takou2025}, which is consistent with the statistical demands of this setting.

Our evaluations reveal that the requisite sample complexity for exact topological reconstruction is predominantly governed by the local topological density, exhibiting sub-linear scaling with respect to the extensive spacetime volume. For instance, while a baseline of $5 \times 10^6$ shots reconstructs the topology of the $d=5, r=5$ surface code without false positives, scaling the memory circuit to $d=5, r=15$ demands only a modest increase to $7 \times 10^6$ shots to maintain the same criterion. Pushing the temporal depth further to $d=5, r=25$ requires only a continued marginal statistical adjustment to $9 \times 10^6$ shots. This mild incremental increase is a natural statistical consequence: while the local variance cascade depth remains topologically invariant, the expanded global candidate pool in deeper circuits marginally elevates the probability of rare statistical tail events crossing the decision boundary, thereby necessitating a slightly tighter suppression of the global noise floor. This near-constant temporal scaling behavior, further detailed in the Appendix, supports the tractability of our framework for deep-circuit quantum hardware.

The computational cost of the clique-search stage follows the same mechanism. As already discussed in the Method section, insufficient statistics generate spurious pair correlations, which in turn induce numerous ghost cliques and enlarge the candidate search space. A dedicated CPU-time benchmark for the $r=5$ rotated surface code across distances $d=5,7,9,11$ is summarized in Table~\ref{tab:cpu_clique_distance}. The data shows that clique search becomes a noticeable bottleneck only in the severely undersampled regime, where fluctuation-induced candidates proliferate rapidly with system size. Once the shot count is sufficiently large, however, the clique-search time becomes very short and increases only relatively slowly with code distance. This confirms that the expensive regime is driven primarily by noise-induced candidate proliferation rather than by code distance alone.

\begin{figure*}[!htb]
    \centering
    \includegraphics[width=0.9\linewidth]{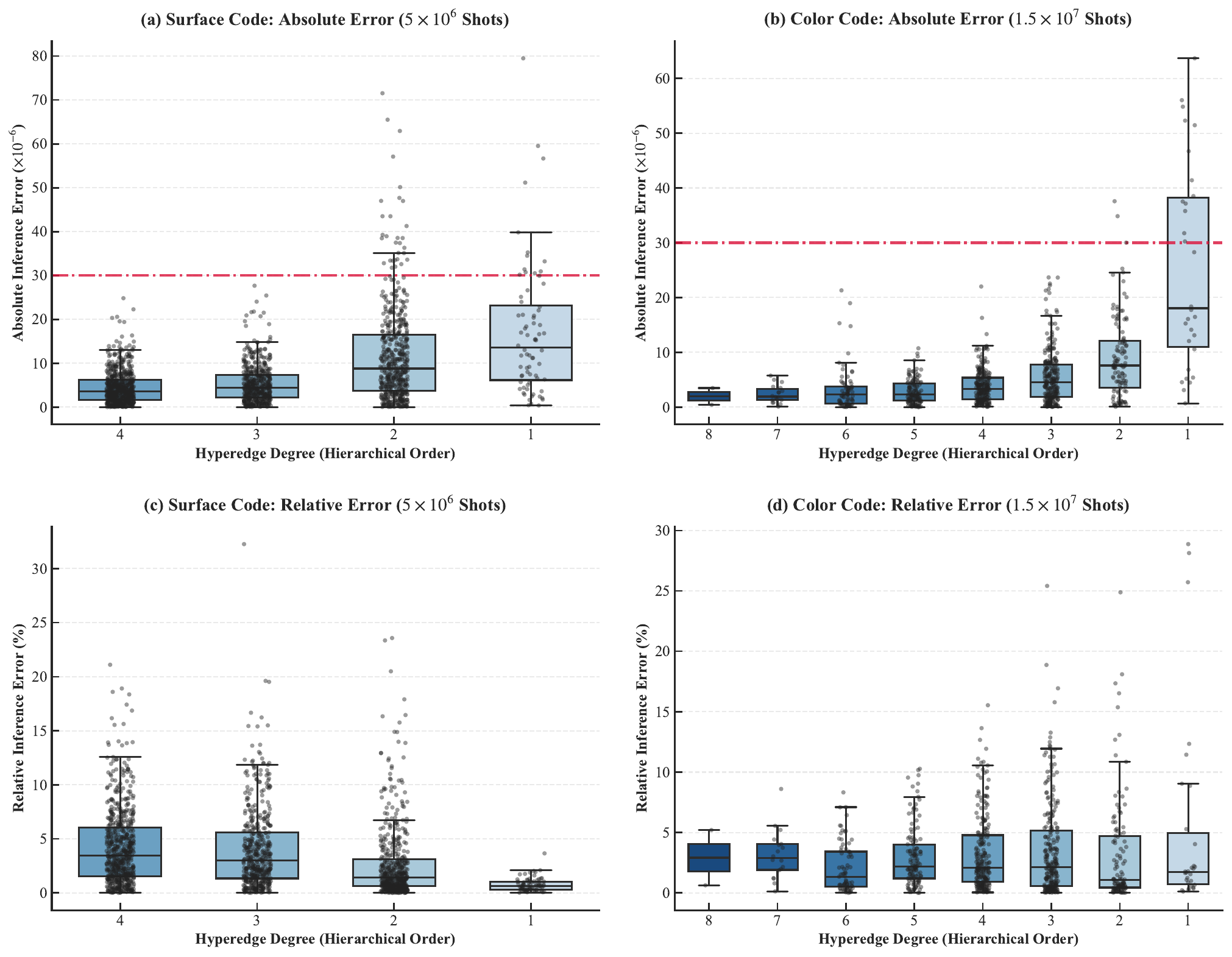}
    \caption{\textbf{Hierarchical accumulation of absolute and relative inference errors (the variance cascade).} Data is shown at the sample sizes required for zero-false-positive topological reconstruction. (a) Surface code ($5 \times 10^6$ shots) and (b) color code ($1.5 \times 10^7$ shots) show a clear inflation of absolute error as hyperedge degree decreases due to algebraic downward subtraction. The horizontal dash-dotted lines denote the multi-body correlation truncation threshold ($\epsilon_{\text{multi}}$). (c) In the surface code, the relative error is strongly suppressed at lower degrees (degrees 1 and 2) because the intrinsic error probabilities in the denominator are comparatively large. (d) In the color code, the large number of high-degree parent hyperedges drives a pronounced variance cascade; consequently, the relative errors at degrees 1 and 2 remain substantially higher than those of the surface code, with outliers exceeding $20\%$.}
    \label{fig:variance_cascade}
\end{figure*}

\subsection{The Variance Cascade and Hierarchical Error Accumulation}
While our inference method serves as a topological prior solver, analyzing the inferred continuous error probabilities reveals a marked statistical divergence driven by the underlying code density. To achieve zero-false-positive hypergraph reconstruction under identical physical error rates and pruning thresholds, the $d=5$ rotated surface code requires $5 \times 10^6$ analysis shots. In contrast, the highly entangled $d=5$ color code demands $1.5 \times 10^7$ shots, three times the statistical volume for surface code. We interpret this sample-complexity disparity through the lens of a \textit{variance cascade}, a phenomenon rooted in our recursive algebraic formulation, as illustrated in Fig.~\ref{fig:variance_cascade}.

Because the exact probability extraction relies on a hierarchical, downward algebraic subtraction, statistical fluctuations originating from finite sampling inherently cascade from parent to child hyperedges. Under a Gaussian approximation for the finite-sample statistics, the absolute variance $\sigma^2$ of a child hyperedge additively absorbs the fluctuations of all its encompassing parent hyperedges, scaling approximately as
\begin{equation}
\sigma_{\text{child}}^2 \approx \sigma_{\text{isolated}}^2 + \sum_{\text{parent}} \sigma_{\text{parent}}^2.
\end{equation}
Governed by this additive accumulation, the observed absolute error for both the surface code (Fig.~\ref{fig:variance_cascade}a) and the color code (Fig.~\ref{fig:variance_cascade}b) visibly inflates as the hyperedge degree decreases. Beyond the physically existing hyperedges, this cascading variance also accumulates onto the analytically inferred probabilities of non-existing candidate edges. Consequently, the densely entangled color code demands a substantially larger statistical ensemble to compress this magnified cascaded noise floor, thereby reliably differentiating unphysical artifacts from genuine multi-body correlations. The same mechanism also increases the chance that analytically inferred probabilities of genuine weak mechanisms become negative at insufficient shot counts, which in turn raises the false-negative rate once physical non-negativity is enforced.

Despite possessing comparable mean absolute errors at lower degrees, the surface code exhibits a substantial population of degree-2 edges whose absolute errors significantly exceed the multi-body noise truncation threshold ($\epsilon_{\text{multi}}$, denoted by the horizontal dash-dotted lines). By comparison, the absolute errors of multi-degree edges in the color code are almost entirely confined below this threshold. Nevertheless, at the shot count used in Fig.~\ref{fig:variance_cascade}a, the surface code still reconstructs the exact fault model. This observation points to a stronger topological robustness in the surface code and, correspondingly, a greater fragility of the color-code error model under statistical fluctuations.

Analyzing the relative error provides a clearer view of this contrast. Although degree-2 edges in the surface code endure large absolute fluctuations, their relative errors remain significantly lower than those in the color code (Fig.~\ref{fig:variance_cascade}c and \ref{fig:variance_cascade}d). This discrepancy is driven by the variance cascade. By the Gaussian approximation, the absolute variance of high-order parent hyperedges cascades additively down to their constituent child edges. Consequently, multi-degree child edges, which possess smaller intrinsic physical error rates ($p_{\alpha}$) in denser topologies, have weaker noise resilience. The large absolute variance inherited from their densely entangled parent hyperedges increases the probability that their analytically inferred error rates are artificially suppressed into the background noise regime or even pushed to negative values. This mechanism indicates that the densely connected color code is more vulnerable to false-negative issues than the surface code, thereby necessitating a larger sample volume to reduce such inferences. This observation is consistent with Table~\ref{tab:master_reconstruction}, wherein, when constrained to $1\%$ and $10\%$ of the ideal sample sizes, the false negatives among the erroneous inferences are markedly higher in the color code than in the surface code. In other words, the variance cascade implies a larger effective variance in the color-code setting and, therefore, a higher probability of negative inferred rates under insufficient statistics; this may also be one reason why direct correlation analysis was not adopted in the Willow color-code experiment \cite{Sivak2025RL}.

\begin{figure*}[!htb]
    \centering
    \includegraphics[width=0.9\linewidth]{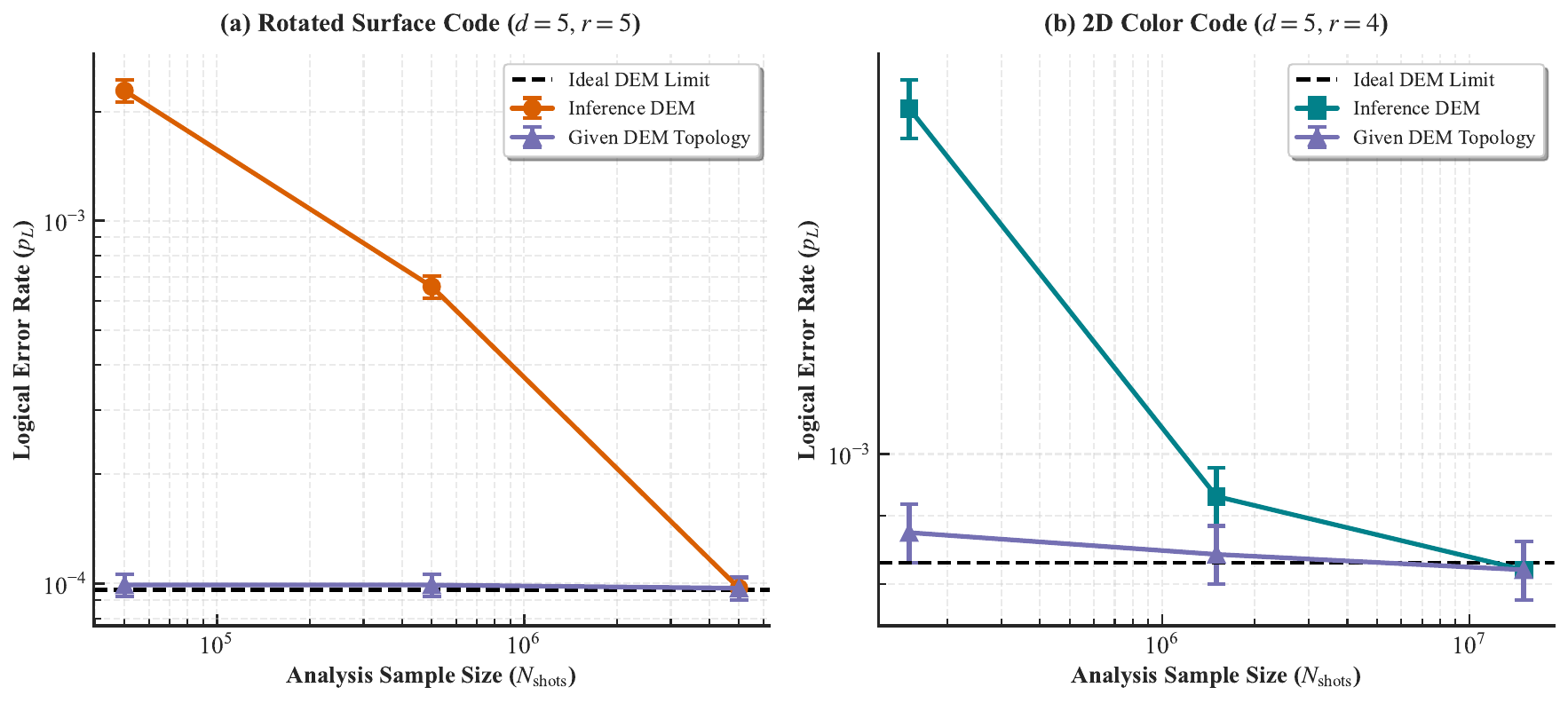}
    \caption{\textbf{End-to-end BP-OSD decoding performance and ablation analysis isolating topological vs. parameter sensitivity.} Logical error rates ($p_L$) are evaluated by collecting 200 (surface code) and 400 (color code) logical errors per point. Data spans three regimes corresponding to $1\%$, $10\%$, and $100\%$ of the statistics required for DEM reconstruction without false positives. For both the (a) surface code and (b) color code, the unpruned Inference DEM fails at low statistics due to dense false-positive short cycles. By contrast, the Given DEM Topology (where the exact ideal topology is fed directly into the global correlation analysis with the pruning threshold set to zero, $\epsilon=0$, solely clipping unphysical negative probabilities) remains much more stable. All curves approach the Ideal DEM limit at the corresponding shot threshold ($5 \times 10^6$ and $1.5 \times 10^7$ shots, respectively), indicating that structural exactness is more important for decoder performance than high-precision parameter estimation in this setting.}
    \label{fig:ler}
\end{figure*}

\subsection{Operational Decoding and the Primacy of Topology}

Subsequently, to evaluate the operational impact of these statistical phenomena, we perform Monte Carlo decoding simulations using Belief Propagation and Ordered Statistics Decoding (BP-OSD)~\cite{Higgott_stimbposd}. To ensure rigorous statistical confidence, we collect 200 logical errors for the surface code and 400 logical errors for the color code. We track decoding performance across three scaling regimes: the required shots for perfect reconstruction ($100\%$), alongside $10\%$ and $1\%$ sparse-sampling baselines.

We employ a crucial ablation control—the ``Given DEM Topology''. In this setup, instead of utilizing the heuristically generated candidate hypergraph, the exact ideal DEM topology is fed directly into the third step of our algorithm (Global Correlation Analysis). To isolate the impact of parameter fluctuations, the global pruning threshold is set to zero ($\epsilon = 0$). This zero-threshold effectively bypasses structural truncation, solely clipping unphysical negative probabilities to zero to prevent the BP message-passing decoder from crashing. Thus, finite statistics are used exclusively to infer the heavily fluctuating continuous error rates over a perfect topological prior.

The decoding results, depicted in Fig.~\ref{fig:ler}, reveal a consistent operational trend across both code families: within the correlation analysis framework, structural exactness (topology) is more important than continuous parameter precision (error rates). For both the surface code (Fig.~\ref{fig:ler}a) and the color code (Fig.~\ref{fig:ler}b), the Inference DEM suffers substantial Logical Error Rate (LER) inflation at low sampling regimes ($1\%$ and $10\%$). This failure is primarily driven by dense false-positive hyperedges injecting many unphysical short cycles into the Tanner graph and degrading the BP message-passing phase.

By contrast, the Given DEM Topology remains comparatively resilient. Even at the minimal $1\%$ sample size—where the inferred parameters suffer from large finite-sample fluctuations and cascaded relative errors frequently exceeding $200\%$—the Given DEM Topology LER stays close to the ideal theoretical baseline. For both codes, supplying the decoder with the exact topology paired with highly distorted, $1\%$-sampled parameters yields a much better LER than utilizing the Inference DEM derived from $10\times$ more data (which still harbors false-positive structural artifacts). Both the Inference DEM and Given DEM Topology LER curves improve monotonically as statistics scale and meet the ideal limit once the $100\%$ shot threshold resolves the exact hypergraph structure.

\section{Discussion}

In this work, we use a Tensor Network Detector Error Model (TNDEM) representation as a structured framework for organizing DEM correlation equations and introduce the Correlation-Analysis-based Hypergraph Reconstruction (CAHR) algorithm for topology discovery, which is a top-down workflow that combines permissive candidate generation with \textit{concurrent} pruning. Unlike sequential greedy growth methods that prematurely truncate signals and frequently conflate statistical noise with genuine physical correlations, this workflow suppresses unphysical artifacts before they propagate downward and bias lower-order constituent edges.

We validated this framework through end-to-end topological reconstruction and BP-OSD decoding simulations on both $d=5$ rotated surface codes and highly dense $d=5$ color codes. Relative to the recent DEM-estimation literature, the surface-code experiments provide a direct comparison point to reconstruction-oriented studies such as \cite{Arms2025EstimatingDE}, while the dense color-code setting addresses a regime where closely matched DEM-reconstruction references remain limited. Our results show that the required sample complexity is governed predominantly by local topological density rather than extensive spacetime volume, with mild growth from $r=5$ to $r=25$. The decoding ablation, comparing the Inference DEM topology against the exact ideal topology, further shows that within this correlation-analysis framework, structural exactness of the topology is operationally more important than high-precision continuous parameter estimation when finite-sample noise is present.

In the preceding experiments, we used $\epsilon \approx p_{\min}/2$ as the working threshold. However, $p_{\min}$ need not be known exactly for the method to remain useful. As shown by the threshold-sensitivity results in Tables~\ref{tab:threshold_surface} and \ref{tab:threshold_color}, varying the cutoff from $p_{\min}/4$ to $3p_{\min}/4$ does not qualitatively change the reconstruction behavior: thresholds in the neighborhood of $p_{\min}$, especially slightly larger ones, remain acceptable in practice. Therefore, in blind experimental settings where $p_{\min}$ is not directly available, a reasonable estimate can still serve as a useful threshold scale. In practice, if reliable circuit and noise-model priors are available, the compiled DEM can provide such an estimate. Otherwise, data-driven procedures such as subsampling-based estimation of the smallest stably resolved nonzero mechanism or conservative noise-floor diagnostics, as used in \cite{Arms2025EstimatingDE}, can serve as fallback strategies.

However, our comparative investigation also exposes a fundamental boundary for purely correlation-analytical inversions. In densely entangled systems like the 2D color code, the downward hierarchical subtraction inevitably drives a pronounced variance cascade, making exact analytical parameter estimation highly susceptible to statistical fluctuations and therefore less suitable as a standalone continuous estimator. This observation clarifies the methodological role of CAHR in a \textit{two-stage decoupled inference paradigm}: CAHR is most naturally used as a Stage-1 topology-discovery tool that resolves the discrete hypergraph and compresses the combinatorial search space, after which Stage-2 methods—such as machine-learning-based decoders, reinforcement learning, or maximum-likelihood estimation—can optimize continuous parameters on the fixed structure.

Finally, we acknowledge the current theoretical boundaries of our framework, which is fundamentally built upon the Pauli error assumption. The present formulation does not provide a controlled, explicit model for general non-Pauli noise, and we have not carried out a quantitative CAHR benchmark under explicit non-Pauli noise models; accordingly, we do not claim quantitative performance conclusions for that setting. Non-Pauli mechanisms—such as leakage~\cite{McEwen2021} or coherent errors~\cite{Darmawan2017Coherent}—therefore require further theoretical generalization for direct modeling within this tensor-network formalism. At the same time, if non-Pauli contributions are present in data without being deliberately separated out, their syndrome-level signatures may still be partially captured by correlation analysis and reflected in the inferred DEM parameters. From the tensor-network perspective, extending TNDEM to faithfully include non-Clifford processes or leakage, and obtaining similar correlation equations, remains an important open problem. Looking ahead, exploring the tensor-network representation for computational and memory optimization, dynamic noise environments, and integration with complementary techniques such as process tomography would further broaden the scope of the framework.

\begin{acknowledgments}
A Python implementation of our algorithm is available at~\cite{code}.
We are grateful to Dong-Yang Feng, Han-Yan Cao and Feng Pan for helpful discussions. 
The work is supported by Projects 12325501 and 12447101 of the National Natural Science Foundation of China.
\end{acknowledgments}

\appendix

\section{Tensor Network Detector Error Model}

The appendix maps the circuit-level DEM into a tensor-network formalism, which we refer to as the Tensor Network Detector Error Model (TNDEM). The purpose of this construction is to provide a structured derivation of the correlation equations used in the main text and to make their connection to tensor-network contraction explicit. While the application of tensor networks to quantum decoding has been explored in contexts such as maximum-likelihood decoding \cite{Piveteau2024}, we use TNDEM here primarily as a derivational and organizational framework for circuit-level noise characterization.

To construct the TNDEM from a general DEM hypergraph $G=(V,E)$, we have two fundamental types of local tensors:
\begin{itemize}
    \item \textbf{Detector Tensors (XOR):} Each detector $d_i \in V$ is represented by a $k$-index XOR tensor, where $k$ is the number of incident error mechanisms. Physically, this tensor enforces the local parity-check constraint, performing a modulo-2 sum (XOR) over all incoming error signals to determine the final classical parity of the detector. Within standard tensor network literature, this is mathematically equivalent to a COPY tensor conjugated by Walsh-Hadamard transforms on all its legs.
    
    \item \textbf{Error Tensors (Diagonal):} Each physical error mechanism $e \in E$ is represented by a diagonal probability tensor. This tensor encodes the binary prior distribution of the fault: the error mechanism triggers all its incident detectors simultaneously with probability $p_e$, and remains dormant with probability $1-p_e$.
\end{itemize}
The complete TNDEM is assembled by contracting the shared indices between the Error Tensors and the Detector Tensors, following the topology of the DEM hypergraph. 

\begin{figure}[!htb]
    \centering
    \includegraphics[width=\linewidth]{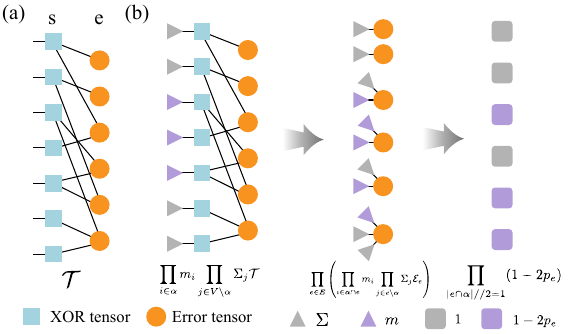}
    \caption{\textbf{Tensor network formulation of the Detector Error Model.} (a) Graphical representation of the TNDEM architecture. Each detector corresponds to an XOR tensor, which enforces a logical parity constraint for propagating Pauli errors. Independent physical error mechanisms are represented by diagonal tensors, parameterized by their fault probability $p_e$ and identity probability $1-p_e$. (b) Visualization of the exact tensor contraction corresponding to Eq.~\eqref{eq:TN_contraction}. By projecting the detector indices onto specific boundary conditions, the global network contraction naturally decouples into a product of independent scalar contributions.}
    \label{fig:TNDEM}
\end{figure}

\section{Derivation of Moment Equations from the TNDEM}

Unlike phenomenological approaches that construct statistical observables heuristically, the analytical power of the TNDEM framework is deeply rooted in its native tensor algebraic properties. In this section, we demonstrate that the exact system of equations governing the hypergraph inference emerges naturally from the fundamental topological rules of the network, specifically, the phase-copying property of XOR tensors. The statistical physics interpretation of these equations is then revealed \textit{a posteriori}.

To extract physical error parameters from experimental syndrome data, one must establish correlation equations linking these microscopic parameters to the expectation values of the macroscopic syndromes. The topological structure of the TNDEM naturally yields a highly structured system of equations amenable to exact analytical inversion. Exploiting the COPY property of the XOR tensor in the dual basis, projecting a detector index onto either the sum state $\Sigma = (1,1)$ or the difference state $m = (1,-1)$ completely factorizes the tensor into a product of independent vectors. Consequently, for an $N$-detector system, enumerating all $2^N$ possible combinations of $\Sigma$ and $m$ boundary projections on the TNDEM yields a complete set of $2^N$ linearly independent equations. Denoting $\alpha \subseteq V$ as the subset of detectors projected onto $m$, with the remaining $V \setminus \alpha$ projected onto $\Sigma$, the network contraction strictly evaluates to:
\begin{equation}\label{eq:TN_contraction}
\begin{aligned}
    \prod_{i\in \alpha} m_{i} \prod_{j\in V\setminus \alpha}\Sigma_j \mathcal{T} 
    =& \prod_{e\in E}\left( \prod_{i\in \alpha\cap e} m_{i} \prod_{j\in e\setminus \alpha}\Sigma_j \mathcal{E}_e \right) \\
    =& \prod_{e\in E} \left[1-p_e + (-1)^{|e \cap \alpha|} p_e\right]\\
    =& \prod_{|e \cap \alpha| \bmod 2 = 1}(1-2p_e).
\end{aligned}
\end{equation}

This multiplicative system of equations can be rigorously linearized by taking the logarithm, rendering it highly tractable. Before proceeding to the analytical solution, we must elucidate the physical significance of contracting the boundary indices with these specific projectors. The projection state $\Sigma$ effectively traces out (marginalizes) the corresponding detector index. Conversely, the projection state $m = (1,-1)$ assigns a statistical weight of $+1$ to the \textit{False} ($0$) syndrome state and $-1$ to the \textit{True} ($1$) syndrome state. This motivates the definition of a spin-like syndrome variable $\sigma_i = 1 - 2s_i$, where $s_i \in \{0,1\}$ is the binary value of detector $d_i$. Under this algebraic mapping, projecting the subset of detectors $\alpha$ onto $m$ mathematically corresponds to computing the physical expectation value of the multi-detector parity observable $\prod_{i\in \alpha} \sigma_{i}$. Denoting this exact multi-body correlation moment as $M_{\alpha} = \langle \prod_{i\in \alpha} \sigma_{i}\rangle$, the tensor contraction rigorous reduces to:
\begin{equation}
    M_{\alpha} = \prod_{|e \cap \alpha| \bmod 2 = 1}(1-2p_e).
\end{equation}
The empty set trivially yields $M_{\emptyset} = \langle1\rangle = 1$. The remaining $2^N -1$ independent equations fully constrain all $2^N -1$ possible hyperedge configurations within the fully connected DEM.

\section{Exact Analytical Solution of the Moment Equations}

In this section, we provide the exact algebraic derivation demonstrating how the helper function $f_{\beta} = \prod_{\alpha\subseteq \beta} M_{\alpha}^{(-1)^{|\alpha|-1}}$ systematically isolates the physical probability of specific multi-body error mechanisms.

Substituting the expression for $M_{\alpha}$ into the definition of $f_{\beta}$ yields:
\begin{equation}
\begin{aligned}
f_{\beta} &= \prod_{\alpha\subseteq \beta}\ \prod_{e\in E, \, |\alpha\cap e| \bmod 2 =1} (1-2p_{e})^{(-1)^{|\alpha|-1}}\\
&= \prod_{e\in E} (1-2p_{e})^{ \sum_{\alpha\subseteq \beta}(|\alpha\cap e| \bmod 2) {(-1)^{|\alpha|-1}}}.
\end{aligned}
\end{equation}
We denote the exponent of the term $(1-2p_e)$ as $C_{\beta,e}$. By replacing the modulo-2 condition with an algebraic indicator function based on powers of $-1$, the exponent can be elegantly rewritten as:
\begin{equation}
\begin{aligned}
    C_{\beta,e} &= \sum_{\alpha\subseteq \beta}(|\alpha\cap e| \bmod 2) {(-1)^{|\alpha|-1}} \\
    &= \sum_{\alpha\subseteq \beta}\cfrac{1- (-1)^{|\alpha\cap e|} }{2} (-1)^{|\alpha\cap e|-1}(-1)^{|\alpha\setminus e|}\\
    &= \sum_{\alpha\subseteq \beta}\cfrac{1- (-1)^{|\alpha\cap e|} }{2}(-1)^{|\alpha\setminus e|}.
\end{aligned}
\end{equation}
This expression can be strategically reshaped by partitioning the subset enumeration into detectors that belong to the error mechanism $e$ and those that do not:
\begin{equation}
    C_{\beta,e} = \sum_{i=0}^{n} \binom{n}{i} \cfrac{1- (-1)^{i} }{2} \sum_{j=0}^{m} \binom{m}{j} (-1)^{j},
\end{equation}
where $n = |\beta \cap e|$ and $m = |\beta \setminus e|$ denote the number of detectors in $\beta$ that are respectively included in $e$ or not. By the binomial theorem, the trailing sum evaluates to $\sum_{j=0}^{m} \binom{m}{j} (-1)^{j} = (1-1)^m$. This summation is identically zero except when $m = 0$, which eliminates the summation entirely. Therefore, a non-zero exponent strictly requires $|\beta \setminus e| = 0$, imposing the geometric condition $\beta \subseteq e$. Consequently, we have $n = |\beta| > 0$. Furthermore, the alternating binomial sum within the remaining part, $\sum_{i=0}^{n} \binom{n}{i}(-1)^{i}$, also evaluates to zero. As a result, the non-zero exponent reduces to:
\begin{equation}
\begin{aligned}
    C_{\beta,e} &= \cfrac{1}{2}\sum_{i=0}^{n} \binom{n}{i}  \\
    &= 2^{n - 1 }.
\end{aligned}
\end{equation}
Remarkably, the resulting exponent depends exclusively on the cardinality of the subset $|\beta|$, entirely independent of the specific geometric structure of $e$. Thus, denoting the exponent simply as $C_{\beta}$, we arrive at the exact analytical inversion formula:
\begin{align}\label{eq:f-p-app}
f_{\beta} &= \prod_{e\in E,\beta\subseteq e} (1-2p_{e})^{C_{\beta}}\\
C_{\beta} &= 2^{|\beta|-1}.
\end{align}

\section{Scalability with Temporal Depth and Constant Sample Complexity}

A critical requirement for any practical fault-tolerant noise characterization protocol is its scalability with respect to the temporal depth (number of QEC rounds, $r$) of the quantum circuit. A naive expectation might assume that as the spacetime volume of the circuit expands, the required sample size for exact topological reconstruction must scale proportionally to suppress the increasing combinatorial space of potential false-positive mechanisms.

However, the Correlation-Analysis-based Hypergraph Reconstruction exhibits favorable sub-linear scaling with respect to the temporal depth of the code. As demonstrated in Table~\ref{tab:round_scaling}, we evaluate the $d=5$ rotated surface code across different temporal depths: $r=5$, $r=15$, and $r=25$. While the baseline $r=5$ circuit achieves zero-false topological reconstruction utilizing $5 \times 10^6$ shots, scaling the memory circuit to $r=25$, which contains nearly six times the total number of true physical hyperedges, requires only a marginal statistical adjustment to $9 \times 10^6$ shots under identical pruning thresholds.

This mild, sub-linear scaling behavior with respect to $r$ is a direct consequence of the localized nature of the \textit{variance cascade}. As established in Section III, statistical variance accumulates strictly downwards from high-degree parent hyperedges to their constituent sub-edges. Extending the number of measurement rounds merely replicates the local extraction circuit topology along the time axis; it increases the extensive volume of the hypergraph but strictly preserves the intensive local topological density (i.e., the maximum hyperedge degree remains 4). Consequently, the maximal length of the hierarchical downward subtraction chain remains constant, rendering the local variance cascade depth topologically invariant with $r$. The slight increase in required statistics (from $5 \times 10^6$ to $9 \times 10^6$ shots) is purely a natural statistical consequence of the extensive scaling: the expanded global candidate pool in deeper circuits marginally elevates the probability of rare statistical tail events crossing the decision boundary, thereby necessitating a slightly tighter suppression of the global noise floor. This confirms that the CAHR is highly scalable and uniquely suited for characterizing the deep-circuit architectures required for practical quantum memory.

\begin{table}[htbp]
\centering
\begin{tabular}{@{}c c c c c@{}}
\toprule
\textbf{Round $r$} & \textbf{Shots} & \textbf{$|E|$} & \textbf{CPU Time (s)} & \textbf{GPU Time (s)} \\
\midrule
5  & $5 \times 10^6$ & 1679 & 8.76 & 4.01  \\
15 & $7 \times 10^6$ & 5759 & 20.07& 14.20 \\
25 & $9 \times 10^6$ & 9839 & 34.29& 31.58 \\
\bottomrule
\end{tabular}
\caption{\textbf{Scaling of sample complexity for perfect reconstruction with temporal depth.} Perfect reconstruction performance on the $d=5$ rotated surface code across varying QEC rounds ($r$). $|E|$ denotes the total number of physical hyperedges, while no false-positive and false-negative appears. Pruning thresholds are fixed at $\epsilon_{\text{single}} = 1.6\times10^{-4}$ and $\epsilon_{\text{multi}} = 3\times10^{-5}$ that $\approx p_{min}/2$ across all depths.}
\label{tab:round_scaling}
\end{table}

\section{Threshold Sensitivity of the Practical Pruning Rule}

To complement the concise discussion in the main text, we provide here the detailed threshold-sensitivity data used to assess the practical robustness of the rule $\epsilon \approx p_{\min}/2$. We evaluate three representative thresholds, $p_{\min}/4$, $p_{\min}/2$, and $3p_{\min}/4$, for both the $d=5,r=5$ rotated surface code and the $d=5,r=4$ 2D color code. The purpose of these tables is not to identify a unique optimal cutoff, but rather to show how the false-positive/false-negative trade-off shifts under finite statistics when the threshold is varied around the practical reference scale. In each table, the two axes correspond to shot count and threshold choice, and each entry is reported as an ordered pair $(\mathrm{FN}, \mathrm{FP})$.

\begin{table}[htbp]
\centering
\small
\begin{tabular}{@{}c c c c@{}}
\toprule
\textbf{Shots} & \textbf{$p_{\min}/4$} & \textbf{$p_{\min}/2$} & \textbf{$3p_{\min}/4$} \\
\midrule
$5\times 10^4$ & $(546, 13270)$ & $(361, 3944)$ & $(323, 2071)$ \\
$5\times 10^5$ & $(3, 1716)$    & $(2, 542)$    & $(30, 170)$   \\
$5\times 10^6$ & $(0, 114)$     & $(0, 0)$      & $(0, 0)$      \\
\bottomrule
\end{tabular}
\caption{\textbf{Threshold sensitivity for the $d=5,r=5$ rotated surface code.} The two axes correspond to shot count and threshold choice, and each entry is reported as $(\mathrm{FN}, \mathrm{FP})$. Lower thresholds suppress FN more aggressively but admit substantially more FP, while higher thresholds reduce FP at the cost of larger FN at intermediate shot counts.}
\label{tab:threshold_surface}
\end{table}

\begin{table}[htbp]
\centering
\small
\begin{tabular}{@{}c c c c@{}}
\toprule
\textbf{Shots} & \textbf{$p_{\min}/4$} & \textbf{$p_{\min}/2$} & \textbf{$3p_{\min}/4$} \\
\midrule
$1.5\times 10^5$  & $(164, 1122)$ & $(156, 600)$ & $(172, 375)$ \\
$1.5\times 10^6$  & $(8, 228)$    & $(11, 47)$   & $(23, 25)$   \\
$1.5\times 10^7$  & $(0, 20)$     & $(0, 0)$     & $(0, 0)$     \\
\bottomrule
\end{tabular}
\caption{\textbf{Threshold sensitivity for the $d=5,r=4$ 2D color code.} The same threshold sweep as in Table~\ref{tab:threshold_surface}, now for the denser color-code benchmark. Each entry is reported as $(\mathrm{FN}, \mathrm{FP})$. The data shows the same qualitative trade-off, but with a sharper sensitivity to finite statistics due to the denser correlated-noise structure.}
\label{tab:threshold_color}
\end{table}

Several trends are immediate from Tables~\ref{tab:threshold_surface} and \ref{tab:threshold_color}. First, $p_{\min}/4$ consistently amplifies false positives in both benchmarks, especially in the low- and intermediate-shot regimes, because the weaker cutoff admits a larger fraction of the finite-sample noise floor. Second, $3p_{\min}/4$ more strongly suppresses false positives, but at intermediate shot counts it more readily truncates genuine weak mechanisms and therefore raises the false-negative count. Third, $p_{\min}/2$ provides the most balanced FN/FP trade-off under finite statistics, which is why we adopt it as the practical threshold in the main text. At sufficiently large shot counts, both $p_{\min}/2$ and $3p_{\min}/4$ reach zero FN and zero FP, whereas $p_{\min}/4$ can still retain residual false positives because its decision boundary remains too close to the finite-sample noise floor.

\section{Clique-Search CPU Time at Different Code Distances}

We next quantify the computational cost of the clique-search stage discussed qualitatively in the Method section. Table~\ref{tab:cpu_clique_distance} reports the CPU core time of the clique-search step for the rotated surface code at $r=5$, across code distances $d=5,7,9,11$ and shot counts $5\times 10^4$, $5\times 10^5$, and $5\times 10^6$. All times are reported in seconds and correspond to the average CPU core time over 5 independent runs on cores of an AMD EPYC 9654 96-Core Processor.

\begin{table}[htbp]
\centering
\small
\begin{tabular}{@{}c c c c c@{}}
\toprule
\textbf{Shots} & \textbf{$d=5$} & \textbf{$d=7$} & \textbf{$d=9$} & \textbf{$d=11$} \\
\midrule
$5\times 10^4$ & 1.011428 & 9.504719 & 53.575279 & 255.009007 \\
$5\times 10^5$ & 0.143893 & 0.216576 & 0.810720  & 2.085380   \\
$5\times 10^6$ & 0.018353 & 0.074927 & 0.146087  & 0.234581   \\
\bottomrule
\end{tabular}
\caption{\textbf{CPU time of clique search for the rotated surface code at different distances.} The table reports the average CPU core time (in seconds) of the clique-search stage over 5 runs, for $r=5$ circuits at different code distances and shot counts. The strong low-shot growth reflects the proliferation of ghost cliques induced by fluctuation-generated pair correlations.}
\label{tab:cpu_clique_distance}
\end{table}

The table supports two main conclusions. First, clique search becomes a noticeable computational bottleneck only when the shot count is severely insufficient. In that regime, fluctuation-induced pair correlations create many ghost cliques, and the candidate search space expands rapidly with system size. This is precisely the quantitative manifestation of the qualitative mechanism already described in the Method section. Second, once the shot count is large enough to suppress most spurious pair correlations, the clique-search time becomes very short and grows only relatively slowly with code distance. 

\bibliographystyle{apsrev4-2}
\bibliography{main}

@book{Nielsen2010,
  author    = {Nielsen, Michael A. and Chuang, Isaac L.},
  title     = {Quantum Computation and Quantum Information},
  publisher = {Cambridge University Press},
  year      = {2010},
  edition   = {10th Anniversary},
  isbn      = {978-1-107-00217-3},
  doi       = {10.1017/CBO9780511976667},
  url = {https://doi.org/10.1017/CBO9780511976667},
  place={Cambridge}
}

@article{Terhal2015,
  title = {Quantum error correction for quantum memories},
  author = {Terhal, Barbara M.},
  journal = {Rev. Mod. Phys.},
  volume = {87},
  issue = {2},
  pages = {307--346},
  numpages = {40},
  year = {2015},
  month = {Apr},
  publisher = {American Physical Society},
  doi = {10.1103/RevModPhys.87.307},
  url = {https://link.aps.org/doi/10.1103/RevModPhys.87.307}
}

@article{Campbell2017,
  title={Roads towards fault-tolerant universal quantum computation},
  author={Campbell, Earl T. and Terhal, Barbara M. and Vuillot, Christophe},
  journal={Nature},
  volume={549},
  number={7671},
  pages={172--179},
  year={2017},
  publisher={Nature Publishing Group},
  doi={10.1038/nature23460},
  url = {https://doi.org/10.1038/nature23460}
}

@article{Spitz2018AdaptiveWE,
author = {Spitz, Stephen T. and Tarasinski, Brian and Beenakker, Carlo W. J. and O'Brien, Thomas E.},
title = {Adaptive Weight Estimator for Quantum Error Correction in a Time-Dependent Environment},
journal = {Advanced Quantum Technologies},
volume = {1},
number = {1},
pages = {1800012},
doi = {https://doi.org/10.1002/qute.201800012},
url = {https://advanced.onlinelibrary.wiley.com/doi/abs/10.1002/qute.201800012},
year = {2018}
}

@article{Gidney2021,
  doi = {10.22331/q-2021-07-06-497},
  url = {https://doi.org/10.22331/q-2021-07-06-497},
  title = {Stim: a fast stabilizer circuit simulator},
  author = {Gidney, Craig},
  journal = {{Quantum}},
  issn = {2521-327X},
  publisher = {{Verein zur F{\"{o}}rderung des Open Access Publizierens in den Quantenwissenschaften}},
  volume = {5},
  pages = {497},
  month = jul,
  year = {2021}
}

@article{Dennis2002,
  title={Topological quantum memory},
  author={Dennis, Eric and Kitaev, Alexei and Landahl, Andrew and Preskill, John},
  journal={Journal of Mathematical Physics},
  volume={43},
  number={9},
  pages={4452--4505},
  year={2002},
  publisher={American Institute of Physics},
  doi={10.1063/1.1499754},
  url = {https://doi.org/10.1063/1.1499754}
}

@article{Fowler2012,
  title = {Surface codes: Towards practical large-scale quantum computation},
  author = {Fowler, Austin G. and Mariantoni, Matteo and Martinis, John M. and Cleland, Andrew N.},
  journal = {Phys. Rev. A},
  volume = {86},
  issue = {3},
  pages = {032324},
  numpages = {48},
  year = {2012},
  month = {Sep},
  publisher = {American Physical Society},
  doi = {10.1103/PhysRevA.86.032324},
  url = {https://link.aps.org/doi/10.1103/PhysRevA.86.032324}
}

@article{Bombin2006,
  title = {Topological Quantum Distillation},
  author = {Bombin, H. and Martin-Delgado, M. A.},
  journal = {Phys. Rev. Lett.},
  volume = {97},
  issue = {18},
  pages = {180501},
  numpages = {4},
  year = {2006},
  month = {Oct},
  publisher = {American Physical Society},
  doi = {10.1103/PhysRevLett.97.180501},
  url = {https://link.aps.org/doi/10.1103/PhysRevLett.97.180501}
}

@article{Harper2020,
  title={Efficient learning of quantum noise},
  author={Harper, Robin and Flammia, Steven T. and Wallman, Joel J.},
  journal={Nature Physics},
  volume={16},
  number={12},
  pages={1184--1188},
  year={2020},
  publisher={Nature Publishing Group},
  doi={10.1038/s41567-020-0992-8},
  url = {https://doi.org/10.1038/s41567-020-0992-8}
}

@article{Takou2025,
  title = {Estimating decoding graphs and hypergraphs of memory quantum error-correction experiments},
  author = {Takou, Evangelia and Brown, Kenneth R.},
  journal = {Phys. Rev. A},
  volume = {112},
  issue = {5},
  pages = {052414},
  numpages = {19},
  year = {2025},
  month = {Nov},
  publisher = {American Physical Society},
  doi = {10.1103/qcz4-nx4r},
  url = {https://link.aps.org/doi/10.1103/qcz4-nx4r}
}

@article{Remm2026Experimentally,
  title = {Experimentally informed decoding of stabilizer codes based on syndrome correlations},
  author = {Remm, Ants and Lacroix, Nathan and B\"odeker, Lukas and Genois, Elie and Hellings, Christoph and Swiadek, Fran\ifmmode \mbox{\c{c}}\else \c{c}\fi{}ois and Norris, Graham J. and Eichler, Christopher and Blais, Alexandre and M\"uller, Markus and Krinner, Sebastian and Wallraff, Andreas},
  journal = {Phys. Rev. Res.},
  volume = {8},
  issue = {1},
  pages = {013044},
  numpages = {21},
  year = {2026},
  month = {Jan},
  publisher = {American Physical Society},
  doi = {10.1103/z1ng-wg3k},
  url = {https://link.aps.org/doi/10.1103/z1ng-wg3k}
}

@article{BlumeKohout2025EstimatingDE,
  title={Estimating detector error models from syndrome data},
  author={Blume-Kohout, Robin and Young, Kevin C.},
  journal={arXiv preprint},
  year={2025},
  eprint={2504.14643},
  archivePrefix={arXiv},
  primaryClass={quant-ph}
}

@article{Arms2025EstimatingDE,
  title={Estimating Detector Error Models on Google's Willow},
  author={Arms, Kregg E. and McHugh, Martin J. and Nyhan, Joseph Edward and Reus, William Frederick and Ulrich, James L.},
  journal={arXiv preprint},
  year={2026},
  eprint={2512.10814},
  archivePrefix={arXiv},
  primaryClass={quant-ph}
}

@article{Panteleev2021,
  doi = {10.22331/q-2021-11-22-585},
  url = {https://doi.org/10.22331/q-2021-11-22-585},
  title = {Degenerate {Q}uantum {LDPC} {C}odes {W}ith {G}ood {F}inite {L}ength {P}erformance},
  author = {Panteleev, Pavel and Kalachev, Gleb},
  journal = {{Quantum}},
  issn = {2521-327X},
  publisher = {{Verein zur F{\"{o}}rderung des Open Access Publizierens in den Quantenwissenschaften}},
  volume = {5},
  pages = {585},
  month = nov,
  year = {2021}
}

@article{Roffe2020,
  title = {Decoding across the quantum low-density parity-check code landscape},
  author = {Roffe, Joschka and White, David R. and Burton, Simon and Campbell, Earl},
  journal = {Phys. Rev. Res.},
  volume = {2},
  issue = {4},
  pages = {043423},
  numpages = {13},
  year = {2020},
  month = {Dec},
  publisher = {American Physical Society},
  doi = {10.1103/PhysRevResearch.2.043423},
  url = {https://link.aps.org/doi/10.1103/PhysRevResearch.2.043423}
}

@article{Wagner2023Learning,
  title = {Learning Logical Pauli Noise in Quantum Error Correction},
  author = {Wagner, Thomas and Kampermann, Hermann and Bru\ss{}, Dagmar and Kliesch, Martin},
  journal = {Phys. Rev. Lett.},
  volume = {130},
  issue = {20},
  pages = {200601},
  numpages = {7},
  year = {2023},
  month = {May},
  publisher = {American Physical Society},
  doi = {10.1103/PhysRevLett.130.200601},
  url = {https://link.aps.org/doi/10.1103/PhysRevLett.130.200601}
}

@article{Sivak2025RL,
  title={Reinforcement Learning Control of Quantum Error Correction}, 
  author={Volodymyr Sivak and Alexis Morvan and Michael Broughton and Rodrigo G. Cortiñas and Johannes Bausch and Andrew W. Senior and Matthew Neeley and Alec Eickbusch and Noah Shutty and Laleh Aghababaie Beni and James S. Spencer and Francisco J. H Heras and Thomas Edlich and Dmitry Abanin and Amira Abbas and Rajeev Acharya and Georg Aigeldinger and Ross Alcaraz and Sayra Alcaraz and Trond I. Andersen and Markus Ansmann and Frank Arute and Kunal Arya and Walt Askew and Nikita Astrakhantsev and Juan Atalaya and Brian Ballard and Joseph C. Bardin and Hector Bates and Andreas Bengtsson and Majid Bigdeli Karimi and Alexander Bilmes and Simon Bilodeau and Felix Borjans and Alexandre Bourassa and Jenna Bovaird and Dylan Bowers and Leon Brill and Peter Brooks and David A. Browne and Brett Buchea and Bob B. Buckley and Tim Burger and Brian Burkett and Nicholas Bushnell and Jamal Busnaina and Anthony Cabrera and Juan Campero and Hung-Shen Chang and Silas Chen and Ben Chiaro and Liang-Ying Chih and Agnetta Y. Cleland and Bryan Cochrane and Matt Cockrell and Josh Cogan and Roberto Collins and Paul Conner and Harold Cook and William Courtney and Alexander L. Crook and Ben Curtin and Martin Damyanov and Sayan Das and Dripto M. Debroy and Sean Demura and Paul Donohoe and Ilya Drozdov and Andrew Dunsworth and Valerie Ehimhen and Aviv Moshe Elbag and Lior Ella and Mahmoud Elzouka and David Enriquez and Catherine Erickson and Vinicius S. Ferreira and Marcos Flores and Leslie Flores Burgos and Ebrahim Forati and Jeremiah Ford and Austin G. Fowler and Brooks Foxen and Masaya Fukami and Alan Wing Lun Fung and Lenny Fuste and Suhas Ganjam and Gonzalo Garcia and Christopher Garrick and Robert Gasca and Helge Gehring and Robert Geiger and Élie Genois and William Giang and Dar Gilboa and James E. Goeders and Edward C. Gonzales and Raja Gosula and Stijn J. de Graaf and Alejandro Grajales Dau and Dietrich Graumann and Joel Grebel and Alex Greene and Jonathan A. Gross and Jose Guerrero and Loïck Le Guevel and Tan Ha and Steve Habegger and Tanner Hadick and Ali Hadjikhani and Michael C. Hamilton and Matthew P. Harrigan and Sean D. Harrington and Jeanne Hartshorn and Stephen Heslin and Paula Heu and Oscar Higgott and Reno Hiltermann and Hsin-Yuan Huang and Mike Hucka and Christopher Hudspeth and Ashley Huff and William J. Huggins and Evan Jeffrey and Shaun Jevons and Zhang Jiang and Xiaoxuan Jin and Chaitali Joshi and Pavol Juhas and Andreas Kabel and Dvir Kafri and Hui Kang and Kiseo Kang and Amir H. Karamlou and Ryan Kaufman and Kostyantyn Kechedzhi and Tanuj Khattar and Mostafa Khezri and Seon Kim and Can M. Knaut and Bryce Kobrin and Fedor Kostritsa and John Mark Kreikebaum and Ryuho Kudo and Ben Kueffler and Arun Kumar and Vladislav D. Kurilovich and Vitali Kutsko and Nathan Lacroix and David Landhuis and Tiano Lange-Dei and Brandon W. Langley and Pavel Laptev and Kim-Ming Lau and Justin Ledford and Joy Lee and Kenny Lee and Brian J. Lester and Wendy Leung and Lily Li and Wing Yan Li and Ming Li and Alexander T. Lill and William P. Livingston and Matthew T. Lloyd and Aditya Locharla and Laura De Lorenzo and Daniel Lundahl and Aaron Lunt and Sid Madhuk and Aniket Maiti and Ashley Maloney and Salvatore Mandrà and Leigh S. Martin and Orion Martin and Eric Mascot and Paul Masih Das and Dmitri Maslov and Melvin Mathews and Cameron Maxfield and Jarrod R. McClean and Matt McEwen and Seneca Meeks and Kevin C. Miao and Zlatko K. Minev and Reza Molavi and Sebastian Molina and Shirin Montazeri and Charles Neill and Michael Newman and Anthony Nguyen and Murray Nguyen and Chia-Hung Ni and Murphy Yuezhen Niu and Logan Oas and Raymond Orosco and Kristoffer Ottosson and Alice Pagano and Agustin Di Paolo and Sherman Peek and David Peterson and Alex Pizzuto and Elias Portoles and Rebecca Potter and Orion Pritchard and Michael Qian and Chris Quintana and Arpit Ranadive and Matthew J. Reagor and Rachel Resnick and David M. Rhodes and Daniel Riley and Gabrielle Roberts and Roberto Rodriguez and Emma Ropes and Lucia B. De Rose and Eliott Rosenberg and Emma Rosenfeld and Dario Rosenstock and Elizabeth Rossi and Pedram Roushan and David A. Rower and Robert Salazar and Kannan Sankaragomathi and Murat Can Sarihan and Kevin J. Satzinger and Max Schaefer and Sebastian Schroeder and Henry F. Schurkus and Aria Shahingohar and Michael J. Shearn and Aaron Shorter and Vladimir Shvarts and Spencer Small and W. Clarke Smith and David A. Sobel and Barrett Spells and Sofia Springer and George Sterling and Jordan Suchard and Aaron Szasz and Alexander Sztein and Madeline Taylor and Jothi Priyanka Thiruraman and Douglas Thor and Dogan Timucin and Eifu Tomita and Alfredo Torres and M. Mert Torunbalci and Hao Tran and Abeer Vaishnav and Justin Vargas and Sergey Vdovichev and Guifre Vidal and Catherine Vollgraff Heidweiller and Meghan Voorhees and Steven Waltman and Jonathan Waltz and Shannon X. Wang and Brayden Ware and James D. Watson and Yonghua Wei and Travis Weidel and Theodore White and Kristi Wong and Bryan W. K. Woo and Christopher J. Wood and Maddy Woodson and Cheng Xing and Z. Jamie Yao and Ping Yeh and Bicheng Ying and Juhwan Yoo and Noureldin Yosri and Elliot Young and Grayson Young and Adam Zalcman and Ran Zhang and Yaxing Zhang and Ningfeng Zhu and Nicholas Zobrist and Zhenjie Zou and Ryan Babbush and Dave Bacon and Sergio Boixo and Yu Chen and Zijun Chen and Michel Devoret and Monica Hansen and Jeremy Hilton and Cody Jones and Julian Kelly and Alexander N. Korotkov and Erik Lucero and Anthony Megrant and Hartmut Neven and William D. Oliver and Ganesh Ramachandran and Vadim Smelyanskiy and Paul V. Klimov},
  journal={arXiv preprint},
  year={2026},
  eprint={2511.08493},
  archivePrefix={arXiv},
  primaryClass={quant-ph},
}

@article{Higgott2025sparseblossom,
  doi = {10.22331/q-2025-01-20-1600},
  url = {https://doi.org/10.22331/q-2025-01-20-1600},
  title = {Sparse {B}lossom: correcting a million errors per core second with minimum-weight matching},
  author = {Higgott, Oscar and Gidney, Craig},
  journal = {{Quantum}},
  issn = {2521-327X},
  publisher = {{Verein zur F{\"{o}}rderung des Open Access Publizierens in den Quantenwissenschaften}},
  volume = {9},
  pages = {1600},
  month = jan,
  year = {2025}
}

@article{mcewen2023relaxing,
  doi = {10.22331/q-2023-11-07-1172},
  url = {https://doi.org/10.22331/q-2023-11-07-1172},
  title = {Relaxing {H}ardware {R}equirements for {S}urface {C}ode {C}ircuits using {T}ime-dynamics},
  author = {McEwen, Matt and Bacon, Dave and Gidney, Craig},
  journal = {{Quantum}},
  issn = {2521-327X},
  publisher = {{Verein zur F{\"{o}}rderung des Open Access Publizierens in den Quantenwissenschaften}},
  volume = {7},
  pages = {1172},
  month = nov,
  year = {2023}
}

@misc{Higgott_stimbposd,
  author       = {Oscar Higgott},
  title        = {stimbposd},
  year         = {2024},
  publisher    = {GitHub},
  journal      = {GitHub repository},
  url          = {https://github.com/oscarhiggott/stimbposd}
}

@article{McEwen2021,
  title={Removing leakage-induced correlated errors in superconducting quantum error correction},
  author={McEwen, M. and Kafri, D. and Chen, Z. and Atalaya, J. and Satzinger, K. J. and Quintana, C. and Klimov, P. V. and Sank, D. and Gidney, C. and Fowler, A. G. and Arute, F. and Arya, K. and Buckley, B. and Burkett, B. and Bushnell, N. and Chiaro, B. and Collins, R. and Demura, S. and Dunsworth, A. and Erickson, C. and Foxen, B. and Giustina, M. and Huang, T. and Hong, S. and Jeffrey, E. and Kim, S. and Kechedzhi, K. and Kostritsa, F. and Laptev, P. and Megrant, A. and Mi, X. and Mutus, J. and Naaman, O. and Neeley, M. and Neill, C. and Niu, M. and Paler, A. and Redd, N. and Roushan, P. and White, T. C. and Yao, J. and Yeh, P. and Zalcman, A. and Chen, Yu and Smelyanskiy, V. N. and Martinis, John M. and Neven, H. and Kelly, J. and Korotkov, A. N. and Petukhov, A. G. and Barends, R.},
  journal={Nature Communications},
  volume={12},
  number={1},
  pages={1761},
  year={2021},
  publisher={Nature Publishing Group UK London},
  doi={10.1038/s41467-021-21982-y},
  url = {https://doi.org/10.1038/s41467-021-21982-y}
}

@article{Darmawan2017Coherent,
  title = {Tensor-Network Simulations of the Surface Code under Realistic Noise},
  author = {Darmawan, Andrew S. and Poulin, David},
  journal = {Phys. Rev. Lett.},
  volume = {119},
  issue = {4},
  pages = {040502},
  numpages = {5},
  year = {2017},
  month = {Jul},
  publisher = {American Physical Society},
  doi = {10.1103/PhysRevLett.119.040502},
  url = {https://link.aps.org/doi/10.1103/PhysRevLett.119.040502}
}

@misc{code,
  author       = {C. Ye}, 
  title        = {{Source code for Correlation-Analysis-based Hypergraph Reconstruction (CAHR)}},
  year         = {2026},
  publisher    = {GitHub},
  journal      = {GitHub repository},
  url          = {https://github.com/huanqingyc/Correlation-analysis-based_HyperDEM_Reconstruction},
}

@article{Piveteau2024,
  title = {Tensor-Network Decoding Beyond 2D},
  author = {Piveteau, Christophe and Chubb, Christopher T. and Renes, Joseph M.},
  journal = {PRX Quantum},
  volume = {5},
  issue = {4},
  pages = {040303},
  numpages = {20},
  year = {2024},
  month = {Oct},
  publisher = {American Physical Society},
  doi = {10.1103/PRXQuantum.5.040303},
  url = {https://link.aps.org/doi/10.1103/PRXQuantum.5.040303}
}

\end{document}